\documentclass[prd,twocolumn,showpacs,superscriptaddress,nofootinbib,floatfix,showkeys,10pt]{revtex4-1}
\usepackage{graphicx}
\usepackage{amsmath}
\usepackage{bm}
\usepackage{yhmath}
\usepackage{mathtools}
\usepackage{wasysym}
\usepackage[colorlinks,citecolor=violet,urlcolor=violet,linkcolor=blue]{hyperref}
\usepackage{subfigure}
\usepackage{color}
\usepackage{cases}
\usepackage{subfigure}
\usepackage{times}
\usepackage{dcolumn,booktabs,bm}
\usepackage{slashed}
\usepackage{amsfonts,amssymb,stmaryrd,latexsym,amsmath}
\usepackage{textcomp}
\usepackage{multirow}
\usepackage{cancel}
\usepackage{array}
\usepackage{orcidlink}

\def\SU3{{\text{SU(3)}_{\rm F}}}

\def \pcs4338{{P_{\psi s}^\Lambda(4338)^0}}

\allowdisplaybreaks

\begin{document}
	
	\title{\textcolor{violet}{Charm content of the proton: An analytic calculation}}
	
	\author{A. R. Olamaei\,\orcidlink{0000-0003-3529-3002}}\email{olamaei@jahromu.ac.ir}
	\affiliation{Department of Physics, Jahrom University, Jahrom, P.~ O.~ Box 74137-66171, Iran}
	\affiliation{School of Particles and Accelerators, Institute for Research in Fundamental 
		Sciences (IPM) P. O. Box 19395-5531, Tehran, Iran}
	
	\author{S.~Rostami\,\orcidlink{0000-0001-7082-6279}}\email{asalrostami.phy@gmail.com}
	\affiliation{Department of Physics, University of Tehran, North Karegar Avenue, Tehran 14395-547, Iran}
	
	\author{K. Azizi\,\orcidlink{0000-0003-3741-2167}}\email{kazem.azizi@ut.ac.ir} \thanks{Corresponding Author}
	\affiliation{Department of Physics, University of Tehran, North Karegar Avenue, Tehran 14395-547, Iran}
	\affiliation{Department of Physics, Do\u gu\c s University, Dudullu-\"Umraniye, 34775 Istanbul, T\"urkiye}
	
	
	\begin{abstract}
According to general understanding, the proton as one of the main ingredients of the nucleus is composed of one down and two up quarks bound together by gluons, described by Quantum Chromodynamics (QCD). In this view, heavy quarks do not contribute to the primary wave function of the proton. Heavy quarks arise in the proton perturbatively by gluon splitting and the probability gradually increases as $Q^2$ increases (extrinsic heavy quarks). In addition, the existence of non-perturbative intrinsic charm quarks in the proton  has also been predicted by QCD. In this picture, the heavy quarks also exist in the proton's wave function. In fact, the wave function has a  five-quark structure $ \vert u u d c \bar{c}\rangle $ in addition to the three-quark bound state $ \vert u u d\rangle $. So far, many studies have been done to confirm or reject this additional component. One of the recent studies has been done by the NNPDF collaboration. They established the existence of an intrinsic charm component at the 3-standard-deviation level in the proton from the structure function measurements. Most of the studies performed to calculate the contribution of the intrinsic charm so far have been based on the global analyses of the experimental data. In this article,  for the first time we directly calculate this contribution by an analytic method. We estimate a $x^{c\bar{c}} = (1.36 \pm 0.67)\% $ contribution for the $ \vert u u d c \bar{c}\rangle $ component of the proton.

	\end{abstract}
	
	\maketitle
	
	\thispagestyle{empty}
	
	\textit{\textbf{\textcolor{violet}{Introduction}}}~~The existence of a non-perturbative intrinsic charm
	quark component in the nucleon
	plays  an increasingly important role in hadron physics.
	Although the structure of the proton is known in the form of a three-quark bound state, QCD   predicts the existence of a non-perturbative intrinsic heavy charm quark contribution to the fundamental structure of the proton. 
	A Fock states of the proton's wave
	function with  a five-quark  structure $ \vert uudc\bar{c}\rangle$  was proposed for the first time by Brodsky, Hoyer, Peterson, and Sakai (BHPS) in Refs. \cite{Brodsky:1980pb,Brodsky:1981se} to explain the large cross-section measured  for the forward open charm production in $ pp $ collisions at the energies
	of the Intersecting Storage Rings (ISR) at CERN \cite{CERN-CollegedeFrance-Heidelberg-Karlsruhe:1978tsl,Giboni:1979rm,Lockman:1979aj,ACCDHW:1979ewo}.
	According to the BHPS model, charm quarks in the nucleon could be either extrinsic or intrinsic. Perturbative extrinsic charm quarks  arise in the proton  where the gluon splits into  charm-anti-charm pairs in the DGLAP $ Q^2 $ evolution and are produced more and more when the $ Q^2 $ scale increases. On the other hand,
	non-perturbative intrinsic charm quarks emerge through the  fluctuations of the nucleon state to the five-quark or virtual meson-baryon states. 
	In recent years, intrinsic charm has been an interesting subject for research from the theoretical, phenomenological, and experimental points of view \cite{Li:2022dnc,Maciula:2022lzk,Li:2022mxp,Hu:2022wsf,Rostami:2016dqi,Aleedaneshvar:2016rkm,Sufian:2020coz}.
	
	In addition to the BHPS model, there have also been some other models to explain the intrinsic charm distribution inside the proton. 
	For instance, in the
	meson cloud model (MCM) which is more dynamical compared 
	to the BHPS, the nucleon can fluctuate to the virtual states composed by 
	a charmed baryon plus a charmed meson \cite{Kumano:1997cy,Thomas:1983fh}. This picture can also be extended to the intrinsic strange content of the nucleon \cite{Goharipour:2018dsa}. 
	The main difference between the BHPS and MCM models is that the charm and anti-charm distributions are different in MCM \cite{Paiva:1996dd}, while they are the same in the BHPS.
	The scalar five-quark model is another approach which was presented by Pumplin \cite{Pumplin:2005yf}. In this
	approach, the distribution for the state $ \vert uudc \bar{c}\rangle$ 
	can be derived from Feynman rules
	(see \cite{Pumplin:2005yf,Hobbs:2013bia} for review of these models).
	
	Although QCD effectively describes the shape of the intrinsic charm distribution, it has nothing to say  quantitatively about the probability of finding the nucleon in the configuration $ \vert uudc\bar{c}\rangle$. However, the experiment can shed light on this issue.
	From the historical point of view, trials to determine the probability of the intrinsic charm content of the proton were started in 1983 when people were motivated to use the BHPS model to explain the data of the European Muon Collaboration (EMC)
	\cite{Hoffmann:1983ah,Harris:1995jx,Steffens:1999hx}. 
	Although these first analyses were not global, they indicated that an intrinsic charm component with probability $ (0.86\pm0.6) \%$ can exist in the nucleon.
	The first global analyses of Parton Distribution Functions (PDFs) considering an intrinsic charm component for the nucleon was performed by the CTEQ collaboration. The aim was also to determine the probability of finding intrinsic charm state in the proton \cite{Pumplin:2007wg,Nadolsky:2008zw,Dulat:2013hea}. 
	Utilizing a wide range of the hard-scattering experimental data, they demonstrated that the charm content can be 2-3 times larger than the value predicted by the BHPS (1$\%$), while this probability was considered to be about $ 0.3 \%$   by the MSTW group \cite{Martin:2009iq}.
	In 2014, the CTEQ collaboration followed their previous works  
	and found a probability of $ 2\% $ for the intrinsic charm considering the BHPS model \cite{Dulat:2013hea}.
	After that, a global analyses was done by Jimenez-Delgado et al \cite{Jimenez-Delgado:2014zga} in which, to show how big the intrinsic charm component could be, they used looser kinematic cuts to include low-$ Q $ and high-$ x $ data. Their prediction was that the intrinsic charm contribution in the proton is about $ 0.5\% $.
	Finally, in the most recent global analyses that has been performed by the NNPDF collaboration \cite{Ball:2022qks}, the intrinsic charm contribution in the flavor content of the proton at the 3-standard-deviation level is estimated to be about $ (0.62 \pm 0.28) \% $.
	
	In addition to these global
	analyses, there have been some  studies to restrict the upper limit on the intrinsic charm content. For instance,  an upper limit  on the intrinsic charm content of the proton, using ATLAS data on measurements of differential cross sections of isolated prompt photons produced in association with a c-jet in $ pp $ collision, is set  in \cite{Bednyakov:2017vck}  as  $ 1.93\% $. According to \cite{Mikhasenko:2012km}, using the ratio of $ \Lambda_{QCD} $ to the difference in the energies of the pentaquark and proton,  the upper bound of the state $ uudc\bar{c}$  is  about $ 1\% $.

	The existence of the intrinsic charm inside  the proton has remarkable and growing experimental support. Several previous and new experiments have been or will be conducted to look for evidences of the intrinsic charm. One of the recent is related to the LHCb data on Z+c-jets over Z+jets  at forward rapidity \cite{LHCb:2021stx}. The data  can be described very well after including a $ 1\% $ intrinsic charm contribution in the proton. Searching for intrinsic charm is also an interesting subject at future experiments like AFTER@LHC \cite{Brodsky:2012vg,Lansberg:2012kf,Lansberg:2013wpx} and ongoing ones.
	The AFTER@LHC experiment is a more suitable laboratory for studying the properties of doubly heavy baryons and it will be interesting to investigate how and to what extent the intrinsic charm affects the results.

	The main goal of this article is to directly calculate, for the first time, the charm component  of proton in a five-quark $ \vert uudc\bar{c}\rangle$ structure analytically using the two-point version of the QCD Sum Rules (QCDSR).
	The remainder of the paper is organized as follows: In the next part we describe the formalism to calculate the proton mass  via the two-point QCDSR. Next, the numerical analyses and results are presented. The final part is devoted to the concluding notes.
	
	\textbf{\textit{\textcolor{violet}{Formalism}}}~~The starting point to calculate any quantity in the QCDSR approach is to write a suitable correlation function (CF). In this case, we write the two-point CF as
	\begin{equation}
		\Pi(q)=i\int d^{4}xe^{iqx}\langle 0|\mathcal{T}\{\eta(x)\bar{\eta}(0)\}|0\rangle,  \label{eq:CorrF1}
	\end{equation}%
	where $\eta (x)$ is the interpolating current of the proton which is the dual of the wave function in the quark model.  $\mathcal{T}$ represents the time-ordering operator and $q$ is the four-momentum of the proton. The time-ordered production of currents is sandwiched between two QCD vacuum states, which corresponds to the creation of the proton in one spacetime point
	(which according to translational invariance can be chosen to be the
	origin) and the annihilation in the spacetime point $x$; after that the result is Fourier transformed to the momentum space.
	
	The interpolating current has two components. The first one corresponds to the ordinary $|uud\rangle$ part of the proton with the spin-parity $ (\frac{1}{2})^+ $:
	\begin{equation}\label{cur.N}
		\eta^{(3q)}(x)=2\varepsilon^{abc}\sum_{\ell=1}^{2}\Big[\Big(u^{T a} (x)CA_{1}^{\ell} d^{b} (x) \Big)A_{2}^{\ell} u^{c} (x)\Big],
	\end{equation}
	where $a$, $b$, $c$ are the color indices and $C$ is the charge conjugation operator.
	The coefficients are $A_{1}^{1}=I$, $A_{1}^{2}=A_{2}^{1}=\gamma_{5}$ and $A_{2}^{2}=\beta$ where $\beta$ is an auxiliary parameter which for the Ioffe current is given by $\beta = -1$. The second part corresponds to the intrinsic charm component in terms of the five-quark structure $|uudc\bar{c}\rangle$. It has a scalar-diquark-scalar-diquark-antiquark type current as:
	\begin{eqnarray}
		\eta^{(5q)}(x)&=&\varepsilon^{ila} \varepsilon^{ijk}\varepsilon^{lmn} \\ \nonumber
		&\times&  u^T_j(x) C\gamma_5 d_k(x)\,u^T_m(x) C\gamma_5 c_n(x)\, \gamma_5 C\bar{c}^{T}_{a}(x) \, ,
	\end{eqnarray}
	where  $i$, $j$, $k$, $\cdots$ are again color indices. This current component has the spin-parity $ (\frac{1}{2})^+ $, as well.
	
	The state of proton including the intrinsic charm component is a superposition of $3q$- and $5q$-components which is:
	\begin{eqnarray}
		|P\rangle = {\cal N}\Big( |uud\rangle + \alpha |uudc\bar{c}\rangle \Big)~,
	\end{eqnarray}
	where $\alpha$ is a  number which indicates the amplitude of the intrinsic charm contribution of the proton and ${\cal N}=(1+ |\alpha|^2)^{-1/2}$ is the normalization constant. Therefore the whole interpolating current  is
	\begin{eqnarray}
		\eta(x) = {\cal N}\Big[\eta^{(3q)}(x) + \alpha ~\frac{\eta^{(5q)}(x)}{m_P^3} \Big].
	\end{eqnarray}
	The factor $m_P^3$ is introduced to ensure that both terms have the same mass dimension. Therefore, in the CF, $\langle 0|\mathcal{T}\{\eta(x)\bar{\eta}(0)\}|0\rangle$ contains four terms where the cross terms $\langle 0|\mathcal{T}\{\eta^{(3q)}(x)\bar{\eta}^{(5q)}(0)\}|0\rangle$ and $\langle 0|\mathcal{T}\{\eta^{(5q)}(x)\bar{\eta}^{(3q)}(0)\}|0\rangle$ give zero contributions according to Wick's theorem for the contraction of the
	quark fields. Therefore, one has:
	\begin{eqnarray}\label{eq:3q5q}
		\Pi(q)&=& i \int d^{4}xe^{iqx} {\cal N}^2 \Big[\langle 0|\mathcal{T}\{\eta^{(3q)}(x)\bar{\eta}^{(3q)}(0)\}|0\rangle \nonumber \\ 
		&& + \frac{|\alpha|^2}{m_P^6} \langle 0|\mathcal{T}\{\eta^{(5q)}(x)\bar{\eta}^{(5q)}(0)\}|0\rangle \Big].
	\end{eqnarray}
	The intrinsic charm contribution which is the probability that the charm component being found in the proton \cite{Brodsky:1980pb} is defined as $x^{c\bar{c}}={\cal N}^2|\alpha|^2$  that we are going to determine.
	
	The only two independent Lorentz structures which can contribute to the CF are $\slashed{q}$ and $U$ and therefore we have
	\begin{eqnarray}\label{str}
		\Pi(q)=\slashed{q}\Pi_1(q^2)+U \Pi_2(q^2),
	\end{eqnarray}
	where the  invariant functions $\Pi_1(q^2)$ and $\Pi_2(q^2)$ have to be calculated. 
	To relate the physical observables like the mass to the QCD calculations, the above mentioned CF has to be calculated in two different regimes. One in terms of hadronic parameters called the physical side which is the real part of the CF and is calculated in the timelike region of the light cone. The other side, called QCD, is evaluated in terms of quarks and gluons using the Operator Product Expansion (OPE)  of the CF. It is the imaginary part of the CF and is calculated in the spacelike region of the light cone.
	These two sides are related to each other via a dispersion integral using the quark-hadron duality assumption which eventually gives us the corresponding sum rule for the mass of the proton. There are contributions from higher states and the continuum which contaminate the ground state. To suppress these and to enhance the ground state contribution we apply the Borel transformation as well as continuum subtraction on both sides of the CF.
	
	The Borel transformation technically enlarges the radius of convergence of the CF integral while leaving the observables unaffected. This transformation along with continuum subtraction introduce two auxiliary parameters, the Borel parameter $M^2$ and the continuum threshold $s_0$, into the calculation, the
	working regions of which have to be determined considering the standard prescriptions of the method.
	
	To calculate the physical side, we insert a complete set of hadronic state with the same
	quantum numbers as the interpolating current of the proton. We
	obtain:
	\begin{eqnarray}\label{PhysSide1}
		\Pi(q)=\frac{\langle 0|\eta(0)|P(q,\lambda)\rangle \langle P(q,\lambda)|
			\bar \eta(0)|0\rangle}{q^2-m_P^2}+...,
	\end{eqnarray}
	where the first term is the isolated ground state proton and dots refer to the higher states and continuum contributions.
	The matrix element $\langle 0|\eta(0)|P(q,\lambda)\rangle $ is determined as 
	\begin{eqnarray}\label{residue}
		\langle 0|\eta (0)|P(q,\lambda)\rangle=\Lambda_P u(q,\lambda),
	\end{eqnarray}
	where $u(q,\lambda)$ is the spinor of the proton with spin $\lambda$ and $\Lambda_P$ is its residue.
	Inserting \eqref{residue} into \eqref{PhysSide1} and summing over the proton's spin one finds the final expression for the physical side of the CF as follows:
	\begin{eqnarray}
		\Pi(q) =\frac{\Lambda^{2}_{P}(\slashed{q}+m_{P} U)}{q^2-
			m_{P}^2}+...,
	\end{eqnarray}
	where the only two independent Lorentz structures $\slashed{q}$ and $U$ emerged as is expected from \eqref{str}.
	
	The QCD side of the CF is calculated in the spacelike sector of the light cone which is the deep Euclidean region. Applying the OPE and using Wick's theorem to contract the quark-antiquark pairs, one can calculate the QCD side of the CF in terms of quark propagators. The
	propagator for the light quarks ($u$ and $d$) reads ~\cite{Yang:1993bp,Reinders:1984sr}:
	\begin{eqnarray}
		&&\mathcal{S}_{q}^{ij}(x)=\frac{i\slashed x}{2\pi ^{2}x^{4}}\delta _{ij}-
		\frac{m_{q}}{4\pi ^{2}x^{2}}\delta _{ij}-\frac{\langle \overline{q}q\rangle
		}{12} ( 1-i\frac{m_{q}}{4}\slashed x) \delta_{ji} \notag \\
		&&-\frac{x^{2}}{192
		}\langle \overline{q}g_{s}\sigma Gq\rangle ( 1-i\frac{m_{q}}{6}\slashed 
		x) \delta_{ij}  \notag \\
		&&-ig_{s}\int_{0}^{1}du\left\{ \frac{\slashed x}{16\pi ^{2}x^{2}}G_{ij}^{\mu
			\nu }(ux)\sigma_{\mu \nu }-\frac{iux_{\mu }}{4\pi ^{2}x^{2}}G_{ij}^{\mu \nu}(ux)\gamma _{\nu } \right. \notag \\
		&&\left. -\frac{im_{q}}{32\pi ^{2}}G_{ij}^{\mu \nu }(ux)\sigma_{\mu \nu }\left[ \ln ( \frac{-x^{2}\Lambda ^{2}}{4}) +2\gamma_{E}\right] \right\} ,  \label{proplight}
	\end{eqnarray}%
	where the subscript $q$ stands for either $u$ or $d$ quark. It is written up to dimension 5, where $\langle \overline{q}q\rangle$ is the quark condensate and $\langle \overline{q}g_{s}\sigma Gq\rangle$ the quark-gluon mixed condensate. The first two terms correspond to the free part and the third and fourth terms to the quark and mixed  condensates, respectively. The integral part represents the one-gluon emission contribution and $\Lambda$ is a cutoff which separates the perturbative and non-perturbative parts. The  propagator of the charm quark can be written as  ~\cite{Yang:1993bp,Reinders:1984sr}:
	\begin{eqnarray}
		&&S_{c}^{ij}(x)=\notag \\
		&&\frac{m_{c}^{2}}{4\pi ^{2}}\frac{K_{1}\left( m_{c}
			\sqrt{-x^{2}}\right) }{\sqrt{-x^{2}}}\delta _{ij} +i\frac{m_{c}^{2}}{4\pi ^{2}
		}\frac{{\slashed x}K_{2}\left( m_{c}\sqrt{-x^{2}}\right) }{\left( \sqrt{%
				-x^{2}}\right) ^{2}}\delta _{ij} \notag \\
		&&-\frac{g_{s}m_{c}}{16\pi ^{2}}\int_{0}^{1}dvG_{ij}^{\mu \nu }(vx) \Big[
		i(\sigma _{\mu \nu }{\slashed x}+{\slashed x}\sigma _{\mu \nu })\frac{%
			K_{1}\left( m_{c}\sqrt{-x^{2}}\right) }{\sqrt{-x^{2}}}  \notag \\
		&& +2\sigma ^{\mu \nu
		}K_{0}\left( m_{c}\sqrt{-x^{2}}\right) \Big] - \frac{\delta_{ij} \langle g_s^2 G^2 \rangle}{576 (2 \pi)^2} \bigg\{ ( i \slashed{x}m_c \notag \\&&- 6) \frac{K_1 (m_c \sqrt{-x^2})}{\sqrt{-x^2}} (-x^2) + m_c(x^2)^2 \frac{K_2(m_c\sqrt{-x^2})}{(\sqrt{-x^2})^2} \bigg\},\notag \\  \label{propc}
	\end{eqnarray}
	where $K_{n}(z)$ is the $n$-th order modified Bessel function of the second kind.
	In \eqref{proplight} and \eqref{propc} we have
	\begin{equation}
		G_{ij}^{\mu \nu }=G_{A}^{\mu \nu }\lambda^{A}_{ij}/2,
	\end{equation}%
	where  $A=1,\,2\,\ldots 8$ and $\lambda^{A}$ are the Gell-Mann matrices.
	Here we use the exponential representation of $K_{n}(z)$ as:
	\begin{eqnarray}
		\frac{K_n(m_c\sqrt{-x^2})}{(\sqrt{-x^2})^n}=\frac{1}{2}
		\int\frac{dt}{t^{n+1}} \exp\left[-\frac{m_c}{2}\left(t-\frac{x^2}{t}\right)\right],
	\end{eqnarray} 
	where
	$-x^2>0$ for the spacelike sector (deep Euclidean region).
	
	In \eqref{str}, the coefficients $\Pi_i(q^2)$ can be written as a dispersion integral as follows:
	\begin{eqnarray}
		\Pi_{i}(q^2)= \int \frac{\rho_{i}(s)}{s-q^2} ds.
	\end{eqnarray}
	Here $\rho_{i}$ are the spectral densities and can be calculated from the imaginary parts of the $ \Pi_{i} $ functions as:
	\begin{eqnarray}
		\rho_{i}(s)=\frac{1}{\pi}Im\Big\{\Pi_{i}(s)\Big\}.
	\end{eqnarray}
	We do not show the very lengthy expressions for the spectral densities in this study.
	According to \eqref{eq:3q5q}, the spectral densities can be decomposed into ordinary and intrinsic charm contributions as: 
	\begin{eqnarray}
		\rho_{i}(s) = {\cal N}^2 \big[\rho_{i}^{(3q)}(s) + \frac{|\alpha|^2}{m_P^6} \rho_{i}^{(5q)}(s)\big].
	\end{eqnarray}

	The last step is to match the coefficients of the corresponding Lorentz structures in \eqref{str}, in both the QCD and physical sides, which gives us the sum rules for the mass and residue of the proton. Considering the ordinary $|uud\rangle$ part only, after applying the Borel transformation and continuum subtraction, as well as using quark hadron duality assumption, one finds the following sum rules:
	\begin{eqnarray}\label{sum1}
		\Lambda^{2}_{P} e^{\frac{-m^{2}_{P}}{M^2}}&=&\int_{s_L}^{s_0} ds \rho_{1}^{(3q)}(s) e^{\frac{-s}{M^2}},\nonumber \\
		\Lambda^{2}_{P} m_{P}e^{\frac{-m^{2}_{P}}{M^2}}&=&
		\int_{s_L}^{s_0} ds \rho_{2}^{(3q)}(s)  e^{\frac{-s}{M^2}},
	\end{eqnarray}
	where $M^2$ and $s_0$ are the auxiliary Borel and continuum threshold parameters for the ordinary $|uud\rangle$ part respectively and $s_L = (2m_u+m_d)^2$.  Then the mass can be calculated from either of the equations \eqref{sum1} (i.e. from either of Lorentz structure $\slashed q$ or $U$) by differentiating the corresponding equation with respect to $z = - \frac{1}{M^2}$ and dividing it over the equation itself which reads:
	\begin{eqnarray}\label{sum2}
		\Big[m^{(3q)}_{P}\Big]^2=\frac{\int_{s_L}^{s_0} ds s\rho^{(3q)}_{i}(s)
			e^{\frac{-s}{M^2}}}{\int_{s_L}^{s_0} ds \rho^{(3q)}_{i}(s)
			e^{\frac{-s}{M^2}}}.
	\end{eqnarray}
   A  similar equation is valid for $m_P^{(5q)}$, the mass of the intrinsic charm contribution $|uudc\bar{c}\rangle$, with corresponding Borel parameter $M'^2$ and continuum threshold $s'_0$.

    To add the intrinsic charm contribution of the proton, following the above recipe, one has to differentiate the sum of $3q$- and $5q$-spectral contributions with respect to $z$. Since the Borel parameter for the $5q$-part is $z' = - \frac{1}{M'^2}$, we need to apply the chain rule 
    \begin{eqnarray}
    	\frac{\partial}{\partial z} = \frac{\partial z'}{\partial z}  \frac{\partial}{\partial z'}~.
    \end{eqnarray} 
	To this, using
	\begin{eqnarray}\label{eq:coef}
		\frac{z}{z'} = \frac{M'^2}{M^2} \simeq \Bigg(\frac{m^{(5q)}}{m^{(3q)}}\Bigg)^2 = b~,
	\end{eqnarray}
	one finds the final sum rule for the proton mass as follows:
	\begin{eqnarray}\label{eq:SRtotal}
		&&m_{P}^2=\\
		&&\frac{\int_{s_L}^{s_0} ds s\rho^{(3q)}_{i}(s)
			e^{\frac{-s}{M^2}} + \frac{|\alpha|^2}{m_P^6} \int_{s_L}^{s'_0} ds (\frac{s}{b})\rho^{(5q)}_{i}(s)
			e^{\frac{-s}{b M^2}}}{\int_{s_L}^{s_0} ds \rho^{(3q)}_{i}(s)
			e^{\frac{-s}{M^2}} + \frac{|\alpha|^2}{m_P^6} \int_{s_L}^{s'_0} ds \rho^{(5q)}_{i}(s)
			e^{\frac{-s}{b M^2}}}\nonumber ~.
	\end{eqnarray}
	
	At low energies, $s_L \leq s \leq s_0$, the charm component behaves as sea quarks and the probability of being observed is considerably low. But at high energies, $s_L \ll s \leq s'_0 $, the charm component emerges as valence-like component and can be detected with an observable probability.
	

	\textbf{\textit{\textcolor{violet}{Numerical Analysis}}}~ The input parameters in the final sum rule \eqref{eq:SRtotal} include quark masses and different quark, gluon and mixed condensates. The condensates are universal non-perturbative parameters, which are determined according to the analyses of many hadronic processes. The  values of these parameters are listed as follows \cite{ParticleDataGroup:2022pth,Belyaev:1982sa,Belyaev:1982cd}:
	\begin{eqnarray}
		&&m_{u}=2.2_{-0.4}^{+0.5}~\mathrm{MeV},\ m_{d}=4.7_{-0.3}^{+0.5}~\mathrm{MeV}, \notag \\
		&&m_{c}=1.27\pm 0.02~\mathrm{GeV}, \notag \\
		&&\langle \overline{q}q\rangle =-(0.24\pm 0.01)^{3}~\mathrm{GeV}^{3}, 
		\notag \\
		&&\langle \overline{q}g_{s}\sigma Gq\rangle =m_{0}^{2}\langle \overline{q}
		q\rangle,\   \\
		&&m_{0}^{2}=(0.8\pm 0.2)~\mathrm{GeV}^{2},  \notag \\
		&&\langle \alpha _{s}G^{2}/ \pi \rangle =(0.012\pm 0.004)~\mathrm{GeV}
		^{4}.\notag
		\label{eq:Parameters}
	\end{eqnarray}
	Moreover, there are four auxiliary parameters ($M^2$, $s_0$, $s'_0$ and $\beta$) that enter the calculations. The working regions of
	these parameters have to be determined. We should calculate their working windows such that the physical quantities  be possibly independent of or have only weak dependence on these parameters. The residual dependencies appear as the uncertainties in the final results.
	
\begin{figure}[h]
		\centerline{\includegraphics[width=0.5\textwidth]{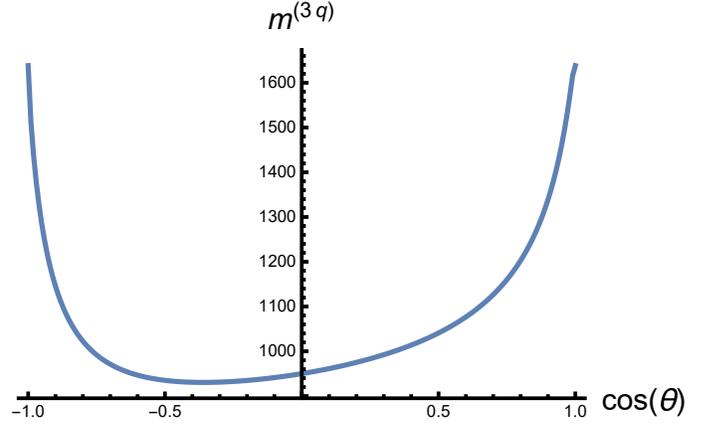}}
		\caption{$m^{(3q)}$ ($ MeV $) as a function of $\cos\theta$ at average values of other auxiliary parameters in their working windows.}
		\label{fig:m3qvst}
	\end{figure}

	The interval for $\beta$ as a mathematical object can be evaluated as follows. Defining a new parameter $\theta$ as $\beta = \tan\theta$ in order to scan $\beta$ in  the whole region from $ -\infty $ to $ +\infty $ by $-1 \leq\cos\theta \leq 1$, and plotting $m^{(3q)}$ and the OPE of the sum rule for the $3q$-part as a function of $\cos\theta$, one can find the interval for $\cos\theta$ where 
	 the physical quantities are more stable and relatively less dependent on it based on the prescriptions of the QCD sum rule method.
	 As an example, we depict the variation of $m^{(3q)}$ with respect to $\cos\theta$ in Fig. \eqref{fig:m3qvst} at average values of other auxiliary parameters.
	From our analyses we find the working region $-0.49 \leq \cos\theta \leq -0.36$ which corresponds to  $-2.60 \leq \beta \leq -1.80$,  in which one can see that $m^{(3q)}$ practically demonstrates  a relatively  good stability with respect to the changes of the  $\cos\theta$. In other words, the variations of $m^{(3q)}$ with respect to $\cos\theta$ is minimal in this working window. As we noted, the residual dependence on the auxiliary parameters contributes to the final error bars of the results. 
	
	To determine the working regions of $M^2$, $s_0$, and $s'_0$, we demand the dominance of Pole Contribution (PC) as well as OPE convergence. They can be quantify by introducing
	\begin{equation}
		\mathrm{PC}^{(j)}=\frac{\Pi^{(j)} (M_{j}^{2},s^j_{0})}{\Pi^{(j)} (M_{j}^{2},\infty )},  \label{eq:PC}
	\end{equation}%
	and
	\begin{eqnarray}
		R^{(j)}(M^{2})=\frac{\Pi ^{\mathrm{Dim-n}(j)}(M_{j}^{2},s^j_{0})}{\Pi^{(j)} (M_{j}^{2},s^j_{0})},
		\label{eq:Convergence}
	\end{eqnarray}%
	where $j$ stands for either  $3q$- or $5q$-component and $\Pi ^{\mathrm{Dim-n}(j)}(M_{j}^{2},s^j_{0})$ is the sum of the three highest dimension operators contributions entered  the OPE expansion of the CF which are 13, 14, 15 for $j=3q$ and 17, 18, 19 for $j=5q$. For the non-perturbative contributions, we follow the principles that \textit{the perturbative part exceeds the total non-perturbative contribution and the higher the dimension of the operator, the lower its contribution to the OPE expansion}.
	Quantitatively, we require that the PC obeys $\mathrm{PC}^{(3q)}\geq 0.5$ which is used to determine $M_{(\mathrm{max})}^{2}$. The lower limit, $M_{(\mathrm{min})}^{2}$, can be found by employing the condition $R^{(3q)}(M_{(\mathrm{min })}^{2})\leq 0.05$ for the sum of last three highest dimensions.
	The above recipes lead to the following windows for the auxiliary parameters: 
	\begin{eqnarray}\label{eq:auxil}
		&&1.15 \leq M^2 \leq 1.50~(\mathrm{GeV}^{2}),~~ 2.15 \leq s_0 \leq 2.30~(\mathrm{GeV}^{2}),\nonumber\\
		&&24 \leq s'_0 \leq 26~(\mathrm{GeV}^{2}).\nonumber\\
	\end{eqnarray}
	\begin{figure}[h]
		\centerline{\includegraphics[width=0.5\textwidth]{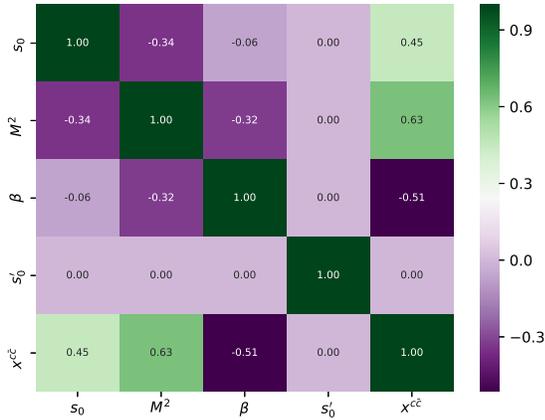}}
		\caption{Symmetric heatmap of the correlation among auxiliary parameters and $x^{c\bar{c}}$. Dark green (dark violet) indicates perfect correlation (anti-correlation). White is the intermediate case of no correlation.}
		\label{heatmap}
	\end{figure}
    
	In order to check the dependence of
	the auxiliary parameters and the physical quantities to each other, we perform a \texttt{python} analysis and plot the resulting heatmap in Fig. \eqref{heatmap}.
	Skipping the last column and line of the figure, it is evident that
	there is a correlation between $M^2$ and $\beta$ and also $M^2$ and $s_0$ (which are related to the dominant $ 3q $ component) as is expected by consistency considerations \cite{Lucha:2007pz}. The other auxiliary parameters are independent from each other. From the last column and line, we see how $ x^{c\bar{c}} $ depends on the auxiliary parameters.
	
	We now proceed to answer the main question of this research work: What is the percentage of the charm content of the proton?
	To this end, considering the working windows of all auxiliary parameters and the values of other inputs, we equate the mass of the proton in \eqref{eq:SRtotal} to the world average of the experimental mass of the proton presented in PDG \cite{ParticleDataGroup:2022pth}. This leads to $x^{c\bar{c}} = (1.36 \pm 0.67)\% $ which, within the uncertainties, overlaps with the estimation of the NNPDF collaboration \cite{Ball:2022qks}. The error presented in the result is due to the uncertainties in the calculations of the windows for the auxiliary parameters as well as errors of other inputs. We should note that $x^{c\bar{c}} $  is the weight of the $ 5q $ component inside the proton and not exactly the $ c \bar{c} $ contribution in our method. To calculate the latter one should develop a method to separate the $ c \bar{c} $ contribution from the $ 5q $ component.  The $ c \bar{c} $ contribution would be the  expectation value $ \langle  P \vert c \bar{c} \vert P\rangle $, which would correspond to the integrated probability of the $ c \bar{c} $ wave function inside the proton in a non-relativistic quark model.

	\textit{\textbf{\textcolor{violet}{Conclusions}}}~~In our
	work, for the first time we used the method of QCD sum rules to determine the contribution of the $ \vert u u d c \bar{c}\rangle $ component in the structure of the proton. 
	Using the two-point correlation function, we found
	$x^{c\bar{c}} = (1.36 \pm 0.67)\% $ for the contribution of this component in the proton. This result, within the presented uncertainties, overlaps with the prediction of the NNPDF collaboration, and is compatible in the upper limit with the  one obtained using the global analysis of the CTEQ collaboration. 
	Our result persuades further dedicated studies of intrinsic charm at future experiments like AFTER@LHC \cite{Brodsky:2012vg,Lansberg:2012kf,Lansberg:2013wpx} and the fixed-target programs of LHCb \cite{LHCb:2018jry}.
	If confirmed experimentally, it provides more insights into the structure  and properties of the proton.
	Moreover, the investigation of atmospheric neutrino measurements presents a viable opportunity to explore the presence of intrinsic charm. As demonstrated in \cite{Laha:2016dri}, such measurements have revealed a noteworthy contribution of intrinsic charm to the atmospheric neutrino flux.
	
	\textit{\textbf{\textcolor{violet}{Acknowledgements}}}~~K. Azizi and S. Rostami are thankful to the Iran Science Elites Federation (Saramadan)
	for the financial support provided under the grant number ISEF/M/99171.
	S. Rostami is grateful to the CERN-TH division for their warm hospitality.
	
	\onecolumngrid

	\twocolumngrid

	\onecolumngrid

\end{document}